\documentclass{ws-ijmpd}
\usepackage[super,compress]{cite}
\begin{document}

\markboth{Chamel}
{ON THE LIE SUBALGEBRA OF KILLING VECTOR FIELDS IN NEWTONIAN SPACE-TIME}

%
\catchline{}{}{}{}{}
%

\title{ON THE LIE SUBALGEBRA OF KILLING-MILNE AND KILLING-CARTAN VECTOR FIELDS IN NEWTONIAN SPACE-TIME}

\author{NICOLAS CHAMEL}

\address{Institut d'Astronomie et d'Astrophysique,
Universit\'e Libre de Bruxelles - CP226, 1050 Brussels,  Belgium\\
nchamel@ulb.ac.be}

\maketitle

\begin{history}
\received{Day Month Year}
\revised{Day Month Year}
\end{history}

\begin{abstract}
The Galilean (and more generally Milne) invariance of Newtonian theory allows for Killing vector fields of a general kind, whereby   
the Lie derivative of a field is not required to vanish but only to be cancellable by some infinitesimal Galilean (respectively Milne) gauge 
transformation. In this paper, it is shown that both the Killing-Milne vector fields, which preserve the background Newtonian space-time structure, 
and the Killing-Cartan vector fields, which in addition preserve the gravitational field, form a Lie subalgebra. 
\end{abstract}


\ccode{PACS numbers: 45.20.D-, 02.40.-k, 04.20.-q}

%

\section{Introduction}

Whereas the concept of Killing vector fields has been widely used in the general relativity theory to derive conservation laws from space-time symmetries 
(see, e.g., Ref.~\refcite{wald84}), its application to the Newtonian context is less well known. This mainly stems from the fact that Newtonian 
mechanics has been traditionally formulated using an ``Aristotelian'' decomposition of space-time, as a direct product of a flat Euclidean three-dimensional 
space with a one-dimensional Euclidean time line. This is only after Einstein proposed his theory of general relativity that a four-dimensional geometric 
formulation of Newtonian theory was developed by Cartan~\cite{cartan23} (see, e.g., Refs.~\refcite{havas64,traut66} for a review). This formulation has been recently 
extended so as to include hydrodynamics~\cite{CK94,CCI,CCII,CCIII} (allowing for fluid and superfluid mixtures), elasticity~\cite{CCC} and elasto-hydrodynamics~\cite{CC06}. 

In Einstein's theory of general relativity, the occurrence of space-time symmetries implies the vanishing of the Lie derivative of the Riemannian metric $g_{\mu\nu}$ (using 
Greek letters $\mu,\nu=0,1,2,3$ for space-time indices) along one or several symmetry generators $k_{\rm a}^\mu$, ${\rm a}=0,1,\dotso$ (see, e.g., Ref.~\refcite{wald84}). 
Although the Lie algebras of the corresponding Killing vector fields in Newtonian space-time have been already studied (see, e.g., Ref.~\refcite{duval09}), the invariance 
of the physical laws of motion under Galilean (and more generally Milne) transformations allows for Killing vector fields of a more general kind, whereby the Lie derivative 
of a field is only required to be cancellable by some infinitesimal Galilean (respectively Milne) gauge transformation~\cite{CCII}. 
Two different kinds of such Killing vector fields were introduced in Ref.~\refcite{CCII}: the Killing-Milne vectors that 
preserve the Milne background structure of Newtonian space-time, and the Killing-Cartan vectors that in addition leave the gravitational 
field invariant. 

In this paper, we shall demonstrate that the Killing-Milne and Killing-Cartan vector fields introduced in Ref.~\refcite{CCII} form a Lie subalgebra after briefly 
reviewing the structure of the Newtonian space-time and discussing the properties of these generalized Killing vector fields. 

\section{Newtonian space-time structure}

Let us first briefly recapitulate the geometric structure of the Newtonian space-time. We shall adopt the same notations as in Ref.~\refcite{CCII}. 
The existence of a universal time $t$ leads to a foliation of the manifold into flat three-dimensional spaces. The pushforward of the three-dimensional 
Euclidean metric $\gamma^{ij}$ (using Roman letters $i,j=1,2,3$ for space indices) yields a symmetric contravariant tensor $\gamma^{\mu\nu}$ in the four-dimensional 
space-time. This tensor itself is not metric since 
\begin{equation}\label{1}
\gamma^{\mu\nu} t_\nu = 0 \, ,
\end{equation}
where $t_\nu =\partial t/\partial x^\nu\equiv\partial_\nu t$. The tensors $\gamma^{\mu\nu}$ and $t_\mu$ specifies the so called Coriolis structure of Newtonian 
space-time. A four-dimensional symmetric covariant tensor $\gamma_{\mu\nu}$ can be 
obtained by pulling back the Euclidean three-dimensional metric $\gamma_{ij}$. The degeneracy condition
\begin{equation}\label{2}
\gamma_{\mu\nu} e^\nu = 0 \, ,
\end{equation}
implies the existence of a so-called ``ether'' frame flow vector $e^\mu$, whose normalization can be chosen such that 
\begin{equation}\label{3}
e^\mu t_\mu = 1 \, .
\end{equation}
The vector $e^\mu$ characterizes a particular Aristotelian coordinate system $\{t,X^i\}$, in which $e^{_0}=1$ and $e^i=0$ corresponding to the usual kind 
of space-time decomposition. The flatness of the three-dimensional hypersurfaces entails the existence of a natural connection, whose components 
$\Gamma_{\mu\nu}^\rho=0$ vanish identically in the corresponding Aristotelian coordinate system (in other words, the covariant derivative $\nabla_\mu$ 
is identifiable with the partial derivative $\partial_\mu$). In an arbitrary coordinate system, some components of the connection may be nonzero. 
However, the associated covariant derivative should satisfy the following conditions: 
\begin{equation}\label{4}
\nabla_\rho\, \gamma^{\mu\nu} = 0 \, , \ \nabla_\rho\, \gamma_{\mu\nu} = 0\, , \ \nabla_\mu t_\nu = 0\, , \ \nabla_\mu\, e^\nu = 0 \, .
\end{equation}
As first shown by Cartan~\cite{cartan23}, the gravitational vector field, defined by 
\begin{equation}\label{5}
g^\mu=-\gamma^{\mu\nu}\nabla_\nu\phi
\end{equation}
$\phi$ denoting the Newtonian gravitational potential, can be absorbed in a gravitationally modified connection 
\begin{equation}\label{6}
\omega_{\mu\nu}^\rho=\Gamma_{\mu\nu}^\rho-t_\mu g^\rho t_\nu\, .
\end{equation}

The vector $e^\mu$ hence the tensor $\gamma_{\mu\nu}$ and the connection $\Gamma_{\mu\nu}^\rho$ are not uniquely defined. The \emph{physical} structure 
of the Newtonian space-time is preserved by Galilean transformations 
\begin{equation}\label{7}
 e^\mu \rightarrow \breve{e}^\mu=e^\mu + b^\mu\, ,
\end{equation}
where $b^\mu$ is a space-like boost velocity vector, whose spatial components $b^i$ in an Aristotelian coordinate system are independent 
of the spatial coordinates $X^i$ and of the time $t$. As first realized by Milne~\cite{milne34}, the Newtonian laws of motion in the presence of gravity are actually 
invariant under a more generic kind of transformations, whereby the spatial components $b^i$ are allowed to depend on $t$, provided that the 
gravitational vector field be transformed as $g^i\rightarrow\breve{g}^i=g^i-a^i$ with $a^i=db^i/dt$. In an arbitrary coordinate system, the boost 
velocity vector will thus be required to satisfy~\cite{CCI} 
\begin{equation}\label{8}
 t_\mu b^\mu=0\, , \ \gamma^{\nu\rho}\nabla_\rho\, b^\mu=0\, . 
\end{equation}
Likewise, the relative acceleration vector field will be given by~\cite{CCI} 
\begin{equation}\label{9}
 a^\mu=e^\nu\nabla_\nu b^\mu\, 
\end{equation}
so that the gravitational vector field transforms as follows:  
\begin{equation}\label{10}
g^\mu\rightarrow\breve{g}^\mu=g^\mu-a^\mu\, .
\end{equation}
The invariance of the Newtonian theory with respect to Milne transformations~(\ref{10}) is embedded in the invariance of the Newton-Cartan connection 
$\breve{\omega}_{\mu\nu}^\rho=\omega_{\mu\nu}^\rho$. 

\section{Killing-Milne and Killing-Cartan vector fields}

The gauge invariance leads to symmetry generators $k^\mu_{\rm a}$ of a new kind~\cite{CCII} such that the corresponding Lie derivative of any (gauge-dependent) 
field $q$ is only required to vanish modulo some infinitesimal gauge transformation $\breve{\rm d}_{\rm a} q$: 
\begin{equation}\label{11}
 \pounds_{\rm a} q + \breve{\rm d}_{\rm a} q = 0\, ,
\end{equation}
where $\pounds_{\rm a} \equiv \vec k_{\rm a} \pounds$ denotes the Lie differentiation operator. Symmetry generators that preserve the background space-time 
structure, namely the tensors $t_\mu$, $\gamma^{\mu\nu}$ and $e^\mu$ (or equivalently $t_\mu$, $\gamma^{\mu\nu}$ and the flat connection $\Gamma_{\mu\nu}^\rho$) 
are termed Killing-Milne vector fields~\cite{CCII}. Since only $e^\mu$ (or equivalently $\Gamma_{\mu\nu}^\rho$) are gauge dependent, the Killing-Milne equations are
\begin{equation}\label{Killing-Milne}
 \pounds_{\rm a} t_\mu = 0 \, , \  \pounds_{\rm a}\gamma^{\mu\nu} = 0\, , \ \pounds_{\rm a} e^\mu + \breve{\rm d}_{\rm a} e^\mu = 0\, .
\end{equation}
These conditions lead to the following equations~\cite{CCII}: 
\begin{equation}\label{Killing-Milne-eq1}
t_\nu \nabla_\mu k_a^\nu = 0\, ,
\end{equation}
\begin{equation}\label{Killing-Milne-eq2}
\gamma^{\rho(\mu} \nabla_\rho k_a^{\nu)} = 0\, ,
\end{equation}
\begin{equation}\label{Killing-Milne-eq3}
e^\nu \nabla_\nu k_a^\mu = b_a^\mu\, ,
\end{equation}
where $b_a^\mu$ is the relevant boost velocity vector field and we have used brackets to indicate index symmetrization. A symmetry generator is termed Killing-Cartan 
vector field~\cite{CCII} if it also preserves the gravitational field $g^\mu$
\begin{equation}\label{Killing-Cartan}
 \pounds_{\rm a} g^\mu + \breve{\rm d}_{\rm a} g^\mu = 0 \, ,
\end{equation}
or equivalently the (gauge-independent) Newton-Cartan connection $\omega_{\mu\nu}^\rho$
\begin{equation}\label{Killing-Cartan-bis}
  \pounds_{\rm a} \omega_{\mu\nu}^\rho = 0\, .
\end{equation}
Such a distinction between Killing-Milne and Killing-Cartan vector fields does not arise in the theory of general relativity since 
the invariance of the Riemannian metric automatically ensures the invariance of the gravitational field. 
The condition~(\ref{Killing-Cartan}) or (\ref{Killing-Cartan-bis}) yields~\cite{CCII}
\begin{equation}\label{Killing-Cartan-eq1}
D_\mu D_\nu k_a^\rho = -R_{\sigma\mu\ \nu}^{\ \ \, \rho} k_a^\sigma\, ,
\end{equation}
or in terms of the gravitational potential  
\begin{equation}\label{Killing-Cartan-eq1-bis}
e^\mu D_\mu \beta_a = - k_a^\mu D_\mu \phi\, ,
\end{equation}
where $D_\mu$ is the Newton-Cartan covariant derivative and $R_{\sigma\mu\ \nu}^{\ \ \, \rho}$ the curvature tensor~\cite{CCI}. 
Let us remark that Eqs.~(\ref{Killing-Milne-eq1}) and (\ref{Killing-Milne-eq2}) can be equivalently expressed as 
\begin{equation}\label{Killing-Cartan-eq2}
t_\nu D_\mu k_a^\nu = 0\, ,
\end{equation}
\begin{equation}\label{Killing-Cartan-eq3}
\gamma^{\rho(\mu} D_\rho k_a^{\nu)} = 0\, .
\end{equation}
This latter equation resembles Killing's equation in Riemannian space-times~\cite{wald84}:   
\begin{equation}\label{Killing-Riemannian}
D_{(\mu} k_{a\,\nu)}=0\, ,
\end{equation}
where $D_\mu$ is the covariant derivative compatible with the metric. 

It immediately follows from (\ref{Killing-Riemannian}) that $k_{a\,\nu}u^\nu$ 
is conserved along the geodesic with tangent vector $u^\nu$~\cite{wald84}. The proof is straightforward: 
\begin{equation}\label{Killing-conservation-Riemaniann}
u^\mu D_{\mu}( k_{a\,\nu}u^\nu)= u^\mu u^\nu D_{\mu}k_{a\,\nu} + k_{a\,\nu} u^\mu D_{\mu} u^\nu = 0\, ;
\end{equation}
the first term vanishes from (\ref{Killing-Riemannian}), and the second from the geodesic equation. In Newtonian space-time, we shall prove that 
the corresponding conserved quantity is 
\begin{equation}
\mathcal{B}_a\equiv k_a^\mu\pi_\mu - \beta_a\, ,
\end{equation}
 where 
\begin{equation}
\pi_\mu \equiv v_\mu - (\frac{1}{2}v^2 +\phi)t_\mu\, , \ v_\mu\equiv\gamma_{\mu\nu} u^\nu\, , \ v^2 \equiv \gamma_{\mu\nu}u^\mu u^\nu = v_\mu u^\mu\, , 
\end{equation}
and $\beta_a$ is the boost potential defined by~\cite{CCI} 
\begin{equation}
 b_a^\mu = \gamma^{\mu\nu}\nabla_\nu \beta_a\, ,
\end{equation}
with $b_a^\mu$ given by Eq.~(\ref{Killing-Milne-eq3}). Let us write the derivative of $\mathcal{B}_a$ along a geodesic with tangent vector $u^\mu$: 
\begin{eqnarray}\label{Killing-conservation}
u^\mu D_{\mu}\mathcal{B}_a=&& u^\mu v_\nu D_{\mu}k_{a}^\nu + k_{a}^\nu u^\mu D_{\mu} v_\nu - \frac{1}{2} t_\nu k_{a}^\nu u^\mu D_\mu v^2 - \frac{1}{2} v^2 t_\nu D_\mu k_{a}^\nu \nonumber \\
&-&t_\nu k_{a}^\nu u^\mu D_\mu \phi - \phi u^\mu t_\nu D_\mu k_{a}^\nu  - u^\mu D_\mu \beta_a\, .
\end{eqnarray}
Although $u^\mu D_{\mu} u^\nu$ vanishes from the geodesic equation, $u^\mu D_{\mu} v_\nu$ does not and is given by 
\begin{equation}\label{27}
u^\mu D_{\mu} v_\nu = u^\mu u^\rho D_{\mu} \gamma_{\nu\rho}=v_\mu g^\mu t_\nu + \gamma_{\mu\nu} g^\mu\, , 
\end{equation}
where we have used Eqs.~(\ref{4}) and (\ref{6}), and the normalization $u^\mu t_\mu=1$. Likewise, we find 
\begin{equation}\label{28}
u^\mu D_{\mu} v^2 = 2 g^\mu v_\mu\, .  
\end{equation}
Using Eqs.~(\ref{27}) and (\ref{28}) as well as (\ref{Killing-Cartan-eq2}) in (\ref{Killing-conservation}), we obtain
\begin{equation}\label{29}
u^\mu D_{\mu}\mathcal{B}_a= u^\mu v_\nu D_{\mu}k_{a}^\nu + k_{a}^\nu g_\nu -t_\nu k_{a}^\nu u^\mu D_\mu \phi  - u^\mu D_\mu \beta_a\, ,
\end{equation}
where $g_\nu\equiv\gamma_{\mu\nu}g^\mu$. Introducing the space-like vector field $v^\mu\equiv u^\mu - e^\mu$, and using Eq.~(\ref{Killing-Milne-eq3}), 
Eq.~(\ref{29}) can be written as 
\begin{equation}
u^\mu D_{\mu}\mathcal{B}_a=  v^\mu v_\nu D_{\mu}k_{a}^\nu + k_{a}^\nu g_\nu -t_\nu k_{a}^\nu e^\mu D_\mu \phi - e^\mu D_\mu \beta_a \, .
\end{equation}
The first term vanishes from the condition~(\ref{Killing-Cartan-eq3}) that qualifies $k_a^\mu$ as a Killing-Milne vector field, and the remaining terms 
cancel each other from Eq.~(\ref{Killing-Cartan-eq1-bis}) that qualifies $k_a^\mu$ as a Killing-Cartan vector field. We have thus proved that $\mathcal{B}_a$
is conserved along a geodesic: 
\begin{equation}
u^\mu D_{\mu}\mathcal{B}_a = 0 \, .
\end{equation}

The maximum number of linearly independent Killing-Cartan vector fields can be determined along the same line of reasoning as in Riemannian space-times~\cite{wald84}. If $k_a^\mu$ and $K_\nu^\mu\equiv D_\nu k_a^\mu$ are known at some point $\mathcal{P}$, $k_a^\mu$ and $K_\nu^\mu$ can be calculated 
at any other point $\mathcal{Q}$ by integrating the following system of equations along any curve connecting  $\mathcal{P}$ and $\mathcal{Q}$
\begin{equation}
\xi^\mu D_\mu k_a^\nu = \xi^\mu K_\mu^\nu\, , \ 
\xi^\mu D_\mu K_\nu^\rho = -\xi^\mu R_{\sigma\mu\ \nu}^{\ \ \, \rho} k_a^\sigma\, ,
\end{equation}
where $\xi^\mu$ is the appropriate vector field and we have used Eq.~(\ref{Killing-Cartan-eq1}). The number of linearly independent Killing-Cartan vector fields is therefore equal to the number of 
initial data, namely the components of $k_a^\mu$ and $K_\nu^\mu$ at point $\mathcal{P}$. Equations~(\ref{Killing-Cartan-eq2}) and (\ref{Killing-Cartan-eq3}), 
which can be expressed as $t_\nu K_\mu^\nu = 0$ and $\gamma^{\rho(\mu}K_\rho^{\nu)}=0$, imply that $K_\mu^\nu$ has only six independent components. With  
the four components of $k_a^\mu$, we can thus conclude that the Newton-Cartan space-time possesses at most ten linearly independent Killing-Cartan vector fields. 

\section{Lie subalgebra of Killing vector fields}

The identity (see, e.g., Ref.~\refcite{choquet82})
\begin{equation} \label{18}
[\vec \xi_{\rm a}, \vec \xi_{\rm b}] \pounds q =  \vec \xi_{\rm a} \pounds \{\vec \xi_{\rm b} \pounds q \}  -   \vec \xi_{\rm b} \pounds \{\vec \xi_{\rm a} \pounds q\}
\end{equation}
for any vector fields $\vec{\xi}_{\rm a}$ and $\vec{\xi}_{\rm b}$ automatically ensures that Killing vector fields of the usual kind 
(whereby the Lie derivative of a field is required to vanish) form a Lie subalgebra with the Lie bracket of two Killing vector fields 
$\vec{k}_{\rm a}$ and $\vec{k}_{\rm b}$ defined by their commutator
\begin{equation}\label{12}
[k_{\rm a}, k_{\rm b}]^\mu = k_{\rm a}^\nu \nabla_\nu k_{\rm b}^\mu - k_{\rm b}^\nu \nabla_\nu k_{\rm a}^\mu=\pounds_{\rm a} k_b^\mu\, ,
\end{equation}
where the last equality follows from the properties of the Lie derivatives (see, e.g., Ref.~\refcite{choquet82}). 
We shall now demonstrate that Killing vector fields of the generic kind form also a Lie subalgebra with the same definition of the Lie bracket. 

Let us consider the successive action of an infinitesimal gauge transformation and the Lie differentiation of a gauge-dependent field $q$ with respect to an 
arbitrary vector field $\vec{\xi}$. Under a change of gauge $e^\mu \rightarrow \breve{e}^\mu=e^\mu+b^\mu$, the ensuing fields $\breve{q}$ will be either a 
function of the boost velocity $b^\mu$ (for $q$ fields like $\gamma_{\mu\nu}$) or of the corresponding acceleration $a^\mu=e^\nu\nabla_\nu b^\mu$ 
(for $q$ fields like $g^\mu$ and $\Gamma_{\mu\nu}^\rho$). We shall treat these two cases separately. For $q$ fields of the first kind, 
an infinitesimal gauge transformation is defined by 
\begin{equation}\label{13}
 \breve{\rm d} q = b^\mu\frac{\partial \breve{q}}{\partial b^\mu}\, ,
\end{equation}
where it is understood that the partial derivative is evaluated in the limit of vanishing boost velocity vector field $b^\mu\rightarrow 0$. 
By a suitable choice of coordinates adapted to the vector field $\vec{\xi}$, the Lie derivative reduces to a partial derivative to some coordinate $x^1$ 
(see, e.g., Ref.~\refcite{wald84}): $\vec{\xi}\pounds=\partial /\partial x^1$. In this coordinate system, the Lie derivative of the field $\breve{\rm d} q$ 
is thus simply given by 
\begin{equation}\label{14}
 \vec{\xi}\pounds \breve{\rm d} q = \frac{\partial b^\mu}{\partial x^1} \frac{\partial \breve{q}}{\partial b^\mu}+ b^\mu \frac{\partial^2 \breve{q}}{\partial x^1 \partial b^\mu} \, . 
\end{equation}
Likewise, the infinitesimal gauge transformation of the field $\vec{\xi}\pounds q$ is given by 
\begin{equation}\label{15}
\breve{\rm d}\{ \vec \xi \pounds  q \} = b^\mu  \frac{\partial^2 \breve{q}}{\partial x^1 \partial b^\mu} \, .
\end{equation}
This shows that the Lie differentiation and the gauge transformation do not commute: 
\begin{equation} \label{16}
[ \vec \xi \pounds ,  \breve{\rm d} ] \,q = \vec \xi \pounds \breve{\rm d} q -  \breve{\rm d}\{ \vec \xi \pounds  q \}= \left(\vec \xi \pounds b^\mu \right) \frac{\partial \breve{q}}{\partial b^\mu} \, .
\end{equation}
Although the field $\breve{q}$ depends on the boost velocity $b^\mu$ corresponding to the specific gauge transformation $e^\mu \rightarrow \breve{e}^\mu=e^\mu+b^\mu$, 
its \emph{functional form}  
$\breve{q}\{b^\mu\}$ is actually gauge-independent. In other words, the fields $\breve{q}_a$ and $\breve{q}_b$ obtained from the same field $q$ by the gauge 
transformations $e^\mu \mapsto \breve{e}^\mu = e^\mu + b_{\rm a}^\mu$ and $e^\mu \mapsto \breve{e}^\mu = e^\mu + b_{\rm b}^\mu$ respectively, are 
such that $\breve{q}_a\{b_a^\mu\}=\breve{q}\{b_a^\mu\}$ and $\breve{q}_b\{b_b^\mu\}=\breve{q}\{b_b^\mu\}$. Consequently, $\partial \breve{q}/\partial b^\mu$
is independent of $b^\mu$. It can thus be seen from Eq.~(\ref{13}) that the commutator~(\ref{16}) represents an infinitesimal gauge transformation of the field 
$q$ with a boost velocity vector field given by $\vec \xi \pounds b^\mu$. 

The two successive gauge transformations  
$e^\mu \mapsto \breve{e}^\mu = e^\mu + b_{\rm a}^\mu$ and $e^\mu \mapsto \breve{e}^\mu = e^\mu + b_{\rm b}^\mu$ are obviously 
equivalent to the gauge transformation $e^\mu \mapsto \breve{e}^\mu = e^\mu + b_{\rm a}^\mu +b_{\rm b}^\mu$. As a consequence, 
the commutator of two infinitesimal gauge transformations vanishes: 
\begin{equation} \label{17}
[ \breve{\rm d}_{\rm a}, \breve{\rm d}_{\rm b} ] \,q = 0 \, .
\end{equation}
Combining Eqs.~(\ref{16}), (\ref{17}) and (\ref{18}) 
we finally find 
\begin{equation} \label{19}
[  \vec \xi_{\rm a} \pounds + \breve{\rm d}_{\rm a},   \vec \xi_{\rm b} \pounds + \breve{\rm d}_{\rm b}] =   \vec \xi_{\rm c} \pounds + \breve{\rm d}_{\rm c}\, ,
\end{equation}
where we have introduced the vector field $\vec \xi_{\rm c} \equiv [\vec \xi_{\rm a}, \vec \xi_{\rm b}]$, and the infinitesimal gauge transformation $\breve{\rm d}_{\rm c}$
is associated with the boost velocity vector field
\begin{equation}\label{20} 
b_{\rm c}^\mu \equiv \vec \xi_{\rm a} \pounds b_{\rm b}^\mu - \vec \xi_{\rm b} \pounds b_{\rm a}^\mu \, .
\end{equation}

A similar analysis can be carried out for $\breve{q}$ fields that depend on $a^\mu$ rather than $b^\mu$. In this case, an infinitesimal gauge transformation is 
defined by 
\begin{equation}
 \breve{\rm d} q = a^\mu\frac{\partial \breve{q}}{\partial a^\mu}\, ,
\end{equation}  
where $a^\mu=e^\nu\nabla_\nu b^\mu$ and the partial derivative is to be evaluated in the limit $a^\mu\rightarrow 0$. The functional form of 
$\breve{q}\{a^\mu\}$ is gauge independent therefore $\partial \breve{q}/\partial a^\mu$ is independent of $a^\mu$. Proceeding as previously, 
equation~(\ref{19}) is found to still hold with the infinitesimal gauge transformation $\breve{\rm d}_{\rm c}$ associated with the boost acceleration vector field
\begin{equation}
a_{\rm c}^\mu \equiv \vec \xi_{\rm a} \pounds a_{\rm b}^\mu - \vec \xi_{\rm b} \pounds a_{\rm a}^\mu \, .
\end{equation}

Let us now consider that $\vec{\xi}_{\rm a}$ and $\vec{\xi}_{\rm b}$ are Killing-Milne or Killing-Cartan vector fields. Using Eqs.~(\ref{Killing-Milne-eq1}), 
(\ref{Killing-Milne-eq2}) and (\ref{Killing-Milne-eq3}), we can show that $a_{\rm c}^\mu=e^\nu\nabla_\nu b_{\rm c}^\mu$. This means that the infinitesimal 
gauge transformation $\breve{d}_{\rm c}$ acting on fields $\breve{q}\{b^\mu\}$ is the \emph{same} as the infinitesimal gauge transformation $\breve{d}_{\rm c}$ 
acting on fields $\breve{q}\{a^\mu\}$. The identity (\ref{19}) thus implies that if $\vec{k}_{\rm a}$ and $\vec{k}_{\rm b}$ are Killing vector fields, their 
commutator $\vec{k}_{\rm c}=[\vec{k}_{\rm a}, \vec{k}_{\rm b}]$ is also a Killing vector field. 
Using Eqs.~(\ref{4}) and (\ref{20}), it can be checked that the boost velocity vector field $b_{\rm c}^\mu$ associated with $\vec{k}_{\rm c}$ is given by   
\begin{equation}\label{21} 
b_{\rm c}^\mu = e^\nu \nabla_\nu [\vec{k}_{\rm a}, \vec{k}_{\rm b}]^\mu =e^\nu \nabla_\nu k_{\rm c}^\mu \, ,
\end{equation}
in accordance with Eq.~(\ref{Killing-Milne-eq3}).

\section*{Acknowledgments}

This work was financially supported by FNRS (Belgium) and the COST Action MP1304.


\end{document}